# Current-in-plane magnetoresistance in chiral-molecule/ferromagnetic metal bilayer due to thermally induced spin polarization

Kouta Kondou[1*], Masanobu Shiga[2], Shoya Sakamoto[2], Hiroyuki Inuzuka[1], Atsuko Nihonyanagi[1], Fumito Araoka[1], Masaki Kobayashi[3,4], Shinji Miwa[2,4,5*], Daigo Miyajima[1], and YoshiChika Otani[1,2,4,5]

[1]*RIKEN, Center for Emergent Matter Science (CEMS), Saitama 351-0198, Japan*
[2]*The Institute for Solid State Physics, The University of Tokyo, Kashiwa, Chiba 277-8581, Japan*
[3]*Department of Electrical Engineering and Information Systems, The University of Tokyo, Bunkyo-ku, Tokyo 113-8656, Japan*
[4]*Center for Spintronics Research Network (CSRN), The University of Tokyo, Bunkyo-ku, Tokyo 113-8656, Japan*
[5]*Trans-scale Quantum Science Institute, The University of Tokyo, Bunkyo, Tokyo 113-0033, Japan*

* Corresponding author: kkondou@riken.jp, miwa@issp.u-tokyo.ac.jp

**We report chirality-induced current-in-plane magnetoresistance (CIP-MR) in chiral molecule/ferromagnetic metal bilayer at room temperature. The previously reported chirality-induced current-perpendicular-to-plane magnetoresistance (CPP-MR) originates from the *chiral induced spin-selectivity* (CISS) effect that needs charge-current passing through the molecule. In contrast, the observed CIP-MR in the present study requires no bias charge current through the molecule. The temperature dependence of CIP-MR suggests thermally induced spin-polarization in the chiral molecules is the key for the observed MR.**

Molecular chirality is an essential factor for inducing spin functionalities in organic materials. Indeed, the spin polarization of electrons passing through chiral molecules has been investigated intensively for two decades[1-13]. This property is called *chiral induced spin selectivity* (CISS) effect, which has been confirmed utilizing various experimental techniques such as photoelectron spectroscopy [2-4,14,15], current-perpendicular-to-plane (CPP) magnetoresistance (MR) measurements [5-7], spin-polarized conductive atomic force microscopy (AFM) [8-10,12,13], inverse spin Hall voltage detection [11] and so on. Surprisingly, large spin polarization comparable to ferromagnets appears at room temperature despite the weak spin-orbit coupling (SOC) of light elements in such chiral molecules [2,3,7-9,14,15]. The physical origins of chirality-induced large spin polarization [16-24] and related phenomena like two-terminal MR [25,26] are still elusive. Nevertheless, such chirality-induced phenomena could apply to nanoscale spin-manipulation for spintronics, quantum computing, and biochemistry.

So far, true and false chiralities are defined by L. D. Barron in 1986 [27]. According to the above, a translating spinning cone is truly chiral while the stationary spinning cone is not. This statement helps understand the chirality-induced phenomena such as nonreciprocal transport in various chiral systems [28,29]. In the case of chirality-induced CPP-MR in chiral molecule/ferromagnet system, spin-polarized bias current passing through the chiral molecule is the translating spinning cone that gives rise to the MR. On the other hand, several experimental studies have recently reported that a chiral molecule could behave like a magnet. For instance, the chiral molecule adsorbed on a superconductor surface exhibits in-gap states similar to the magnetic impurity state in tunneling spectra [30]. Moreover, the chirality-dependent effective magnetic field is generated in the ferromagnetic metal/chiral molecule bilayer [31,32]. Remarkably, no bias charge current flows through the molecule, implying that the magnetization, *i.e.,* spontaneous spin polarization, might emerge in the chiral molecules. However, the existence of the chirality-dependent spontaneous spin polarization is not allowed because the spin polarization itself is falsely chiral [27]. Thus, to explain the spin polarization indicated by the previous studies [30-32], one should consider a thermally driven microscopic flow of spin angular momentum to make the system truly chiral.

This paper reports the first observation of the current-in-plane magnetoresistance (CIP-MR) effect at room temperature in a chiral molecule/ferromagnet bilayer system. In general, the MR effect appears in ferromagnetic metal/non-magnetic metal/ferromagnetic metal multilayer systems. Figure 1(a) shows the schematic illustration of the CIP-GMR effect. Depending on the relative magnetization directions in the two ferromagnetic layers, overall electrical resistance varies with the change in spin-dependent electron scattering due to the shift in chemical potential at the interface [33-36]. Similarly, we investigated the chirality-induced CIP-MR effect using the chiral molecule/ferromagnet heterostructure shown in Figure 1(b). The CIP-MR in Fig. 1(b) reflects spontaneous spin polarization emerging in the chiral molecule via spin-dependent scattering. Note that the device configuration in this paper is essentially different from the previously reported CPP-MR, where bias charge current flows through the chiral molecules along their helical axis [5,7].

In the present study, we employ the chiral molecules of *Plus* (*P*)- and *Minus* (*M*)-PbPc-DTBPh, as shown in Figure 1(c), which have respectively *right-* and *left-*handed helicities. The detailed methods for the synthesis are reported elsewhere [32]. Figure 1(d) shows the scanning electron microscope image and a schematic illustration of the measured multilayer, Ni(5 nm)/(*P*)- or (*M*)-PbPc-DTBPh(0-1 nm)/MgO(2 nm)/AlO$_x$(5 nm) multilayer on a thermally oxidized Si substrate. The PbPc-DTBPh layer was deposited at an evaporation rate of $5 \times 10^{-3}$ nm s$^{-1}$ (0.05 Å s$^{-1}$) in an ultrahigh vacuum ($< 5 \times 10^{-7}$ Pa). The molecule thickness was determined using a quartz thickness monitor and atomic force microscopy. One molecular layer of the PbPc-DTBPh is approximately 0.6 nm. The Ni and oxide capping layers of MgO/AlO$_x$ were grown at an evaporation rate of $1 \times 10^{-2}$ nm s$^{-1}$ (0.1 Å s$^{-1}$) using an electron beam deposition in an ultra-high vacuum ($< 3\times 10^{-7}$ Pa). We should note that the molecular orientation does not affect the molecular chirality of PbPc-DTBPh because of its structurally determined nature [7,32]. The Ni/PbPc-DTBPh multilayer films were patterned into 5 μm × 100 μm rectangular shapes using the photolithography and Ar ion milling process. Sample resistance was measured by the conventional 4 terminal method to remove the contact resistance. In this measurement, most of the probe currents flow in the Ni layer, whose conductivity is about 15 orders of magnitude larger than the molecule layer.

Figure 1(e) shows the X-ray photoemission spectroscopy (XPS) spectrum of the Ni 2*p* core level in Ni/PbPc-DTBPh measured at room temperature. Comparing with the control sample of Ni/MgO, we find a charge transfer (CT) that Ni is doped with an electron from the PbPc-DTBPh regardless of its chirality. It implies that a hybridized interface state (HIS) appears between the ferromagnet conduction band and HOMO of the chiral molecule [37-39] since there is a high and narrow density of state (DOS) peak in the vicinity of the Fermi level of Ni.

Figure 2 shows the typical MR curve in Ni/(*P*)-PbPc-DTBPh, Ni, and Ni/(*M*)-PbPc-DTBPh-based multilayer, respectively. The Ni sample without the PbPc-DTBPh molecule is the control sample. The overall sample resistance in Ni/PbPc-DTBPh bilayer devices is slightly larger than the control sample. The direction of the applied external magnetic field ($H_z$) is normal to the film plane. The applied direct current for resistance measurement is 50 μA. These measurements were carried out at room temperature. All of the sample resistances in Figures 2(a)-(c) show the typical anisotropic MR effect of the Ni layer as the Ni layer magnetization direction gradually rotates from in-plane to out-of-plane with $H_z$ field. The MR values at positive and negative saturation magnetic fields are almost the same in the control sample. Thus, the surface roughness-induced GMR effect in the Ni layer can be negligibly small in the present sample. In the Ni/(*P*)-PbPc-DTBPh bilayer sample, the MR value at the positive saturation magnetic field is *smaller* than at the negative one, as shown in Figure 2(a). Most interestingly, in the case of the opposite helicity in the molecule, i.e., Ni/(*M*)-PbPc-DTBPh bilayer, shown in Figure 2(c), the tendency is reversed; the value at the positive saturation magnetic field is *larger* than that at the negative one. The MR change is independent of the magnetic field sweep direction and the probe current amplitude from -100 to +100 µA (Supplemental Information), implying that the MR change is not related to the hysteretic magnetic domain formation in the Ni layer and the Joule heating by the applied current.

Hence, this must be a chirality-induced CIP-MR effect. The results obtained in the absence of the current flow in the chiral molecular layer strongly suggest that the chirality-dependent spin polarization emerges inside the molecules, as shown by red and blue arrows in the schematic illustrations above Figs. 2(a) and (c). These results suggest CIP-MR may originate from the spin-dependent scattering at Ni/chiral molecule interface. The above facts are also supported by the recent experimental report of the chirality-induced effective magnetic field in the Fe/PbPc-DTBPh/MgO multilayer [32]. The spin polarization induced in the PbPc-DTBPh corresponds to the exchange-coupled magnetization direction observed in the Fe/PbPc-DTBPh/MgO trilayer.

Figure 2(d) shows the MR ratio $\Delta R(0.8\,\text{T})/R(0\text{T})\times 100$ (%) with $\Delta R(0.8\,\text{T}) = R(+0.8\,\text{T}) - R(-0.8\,\text{T})$ as a function of the molecular thickness up to 1 nm. $R(\pm 0.8\,\text{T})$ and $R(0\,\text{T})$ are sample resistance at $H_z = \pm 0.8$ T and 0 T, respectively. The blue and red squares in Figure 2(d) respectively correspond to the MR ratios for (*M*)-PbPc-DTBPh and (*P*)-PbPc-DTBPh. With an increase in the amount of molecular adsorption on Ni layer surface up to ~0.6 nm (~one molecular layer: 1 ML), the values of $\Delta R$ for both (*M*)- and (*P*)-PbPc-DTBPh increase gradually. This result supports that the molecular adsorption on the ferromagnetic layer surface causes the MR effect. On the other hand, above ~0.6 nm (1 ML), the values of $\Delta R$ for (*M*)- and (*P*)- molecules considerably decrease with PbPc-DTBPh thickness. This trend is closely related to the Rashba interface formation, increasing the adsorbent achiral PbPc molecules on a Cu surface [40]. In Ref. 38, the deformed PbPc stacking caused a significant decline of spin-to-charge conversion signal above ~1 ML of PbPc on the Cu surface. The XPS spectra shown in Fig. 1(e) indicate a CT between the chiral molecules and Ni layers. Therefore the observed molecular thickness dependence suggests that HIS at Ni/PbPc-DTBPh interface plays a crucial role in the emergence of the chirality-induced CIP-MR. The difference in MR ratios for (P)-PbPc-PbPc-DTBPh and (M)-PbPc-PbPc-DTBPh could originate from a slight inhomogeneity of the molecular film.

Here, we discuss the mechanism of the chirality-induced CIP-MR effect, i.e., how the spontaneous spin polarization can appear in the chiral molecule. Figure 3(a) shows the resistance change $\Delta R$ at various temperatures from 300 K to 10 K for Ni/(*M*)-PbPc-DTBPh (0.6 nm~1 ML) bilayer device. To eliminate the anisotropic MR contribution of the Ni layer, we plot the resistance change $\Delta R$ defined as $\Delta R(\pm H_z) = R(\pm H_z) - R(\mp H_z)$, where $R(\pm H_z)$ are sample resistances at the positive and negative saturation magnetic fields, respectively.

The $\Delta R$ saturates around ±0.5 T, which corresponds to the saturation field of the Ni layer. A decrease in temperature from 300 K diminishes the effect, making $\Delta R$ almost zero below about 50 K. Similarly, the MR ratio, given by $\Delta R(1\,\text{T})/R(0\text{T}) \times 100$ (%), in Figure 3(b) exhibits a gradual decrease and takes almost zero below about 50 K, suggesting thermally induced spin polarization of the chiral molecules on Ni layer. By fitting the data to Boltzmann distribution function $\exp(-\Delta/k_B T)$, where $\Delta$, $k_B$, and $T$ are barrier height, Boltzman constant, and temperature, $\Delta$ can be estimated to be ~10 meV. Its small energy difference may indicate the existence of HIS at Ni/PbPc-DTBPh interface.

      A possible mechanism to generate the thermally induced spin polarization in the chiral molecule is schematically shown in Figure 4. Thermally excited hopping charge transport between a conduction band of metal and HOMO of the chiral molecule may occur; the hopping rate increases with temperature. When an electron moves from metal to chiral molecule, the electron can be spin-polarized along out-of-plane (+$z$-direction), assuming that the CISS or current-induced spin polarization effect [41-43] occurs during the electron transport in HIS, as shown in the red arrow. Simultaneously, an electron moves from chiral molecule to metal with opposite spin-polarization. We assume that spin relaxation time in the molecule $\tau_S^{mol}$ is longer than transit time $\tau_S^t$ between metal and chiral molecule. Here, the spin relaxation time in metal $\tau_S^{met}$ [44] is several order smaller than $\tau_S^{mol}$ [45-47], and the net spin polarization in the system would remain inside the chiral molecule. In this mechanism, the metal layer does not need to be ferromagnet to generate the spin polarization in the chiral molecule. The finite CIP-MR was observed even in the sample where the thin Cu layer is inserted between Ni and PbPc-DTBPh (Supplemental information). This thermally induced spin-polarization can explain the temperature dependence of the chirality-induced CIP-MR effect shown in Figure 3(b). The previous works on the temperature dependence of two-terminal CPP-MR have shown that the MR ratio diminished with decreasing the temperature [5,9]. Thus, the thermally induced spin-polarization may cause a declined MR ratio in the chiral molecule.

      In summary, we demonstrated the chiral-molecule-induced CIP-MR. In contrast to the CPP-MR due to the conventional CISS effect, the MR does not require the bias charge current in the chiral molecule. These results imply the emergence of spin polarization inside the chiral molecule. We believe that thermally induced spin polarization due to spin selective transport and different spin relaxation time between the metal and chiral molecules is crucial for the emergence of the MR effect. Our findings on magnetoresistance could trigger the dramatic development of molecular spintronics.

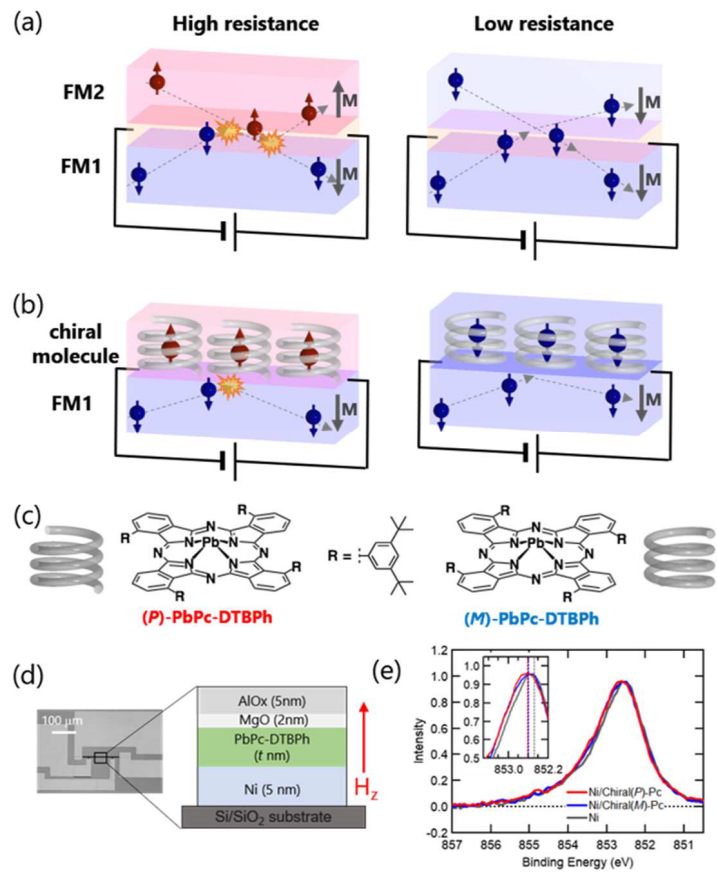

**Figure 1 Concept of chiral-molecule induced current-in-plane giant magnetoresistance.** (a), (b) A schematic illustration of conventional metal-based current-in-plane giant magnetoresistance (CIP-GMR) effect and chiral-molecule induced MR effect. Springs beside each chiral molecule denote the helicity. (c) Structure of the chiral molecule. Right-handed helicity: (*P*)-PbPc-DTBPh and left-handed helicity: (*M*)-PbPc-DTBPh. (d) Scanning electron micrograph of a measurement sample and schematic of the multilayer (e) XPS spectrum in Ni/PbPc-DTBPh films.

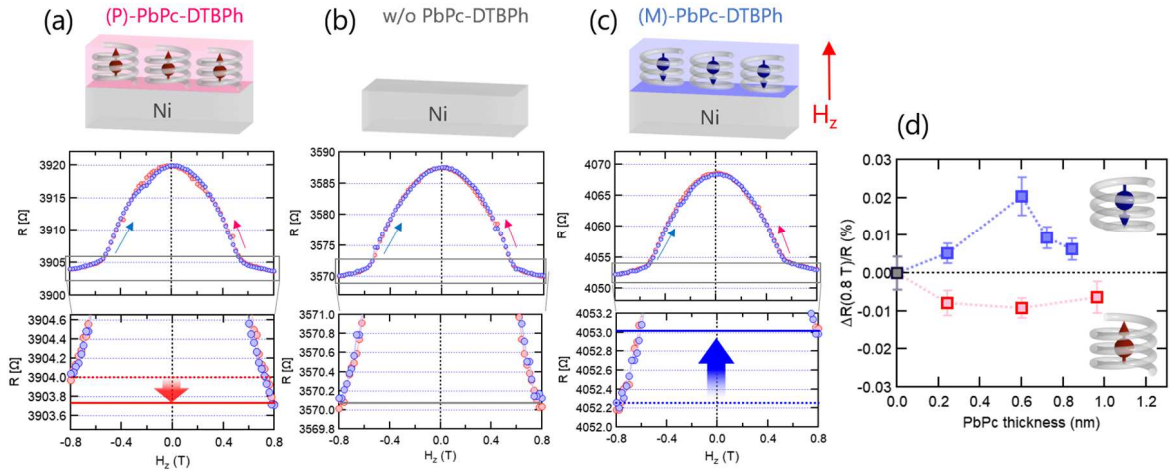

**Figure 2 Molecular chirality and thickness dependence.** Magnetoresistance measurement in (a) Ni/(P)-PbPc-DTBPh film, (b) Ni film as a control sample, (c) Ni/(M)-PbPc-PbPc-DTBPh film. Blue and pink open plots correspond to the experimental data in each magnetic field sweep direction shown by red and blue arrows. (d) Molecular thickness dependence of magnetoresistance ratio ΔR/R due to chiral molecule induced CIP-MR effect at room temperature. Plots at left-half, zero and right half is correspond to the data in (*P*)-PbPc-PbPc-DTBPh, control sample and (*M*)-PbPc-DTBPh, respectively.

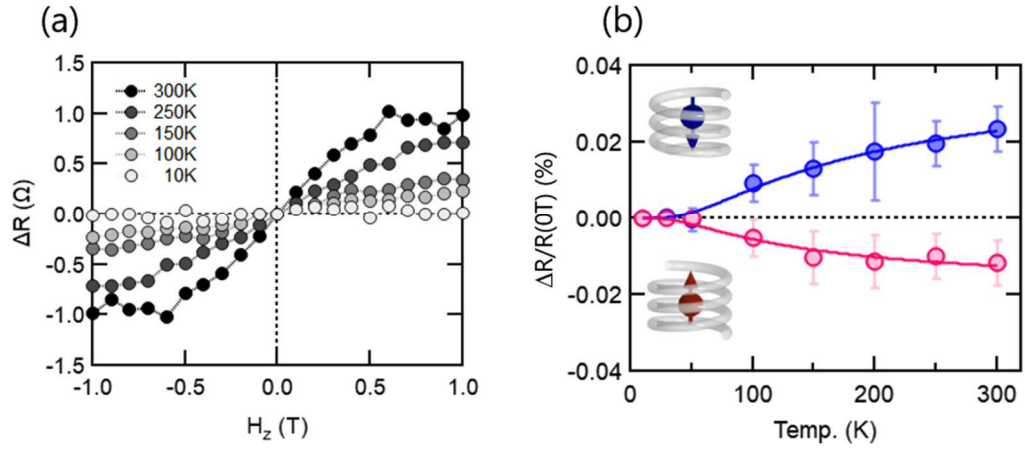

**Figure 3 Temperature dependence.** (a) Variation of chirality-induced CIP-MR effect ΔR in Ni/(M)-PbPc-PbPc-DTBPh (0.6 nm) film as a function of the applied magnetic field at various temperatures. Applied direct current is 10 μA (b) ΔR/R as a function of measurement temperature, in which Blue and Red plots correspond to data for Ni/(*M*)-PbPc-PbPc-DTBPh and Ni/(*P*)-PbPc-PbPc-DTBPh, respectively. Bule and Red curves are the fitting curves by the Boltzmann distribution function.

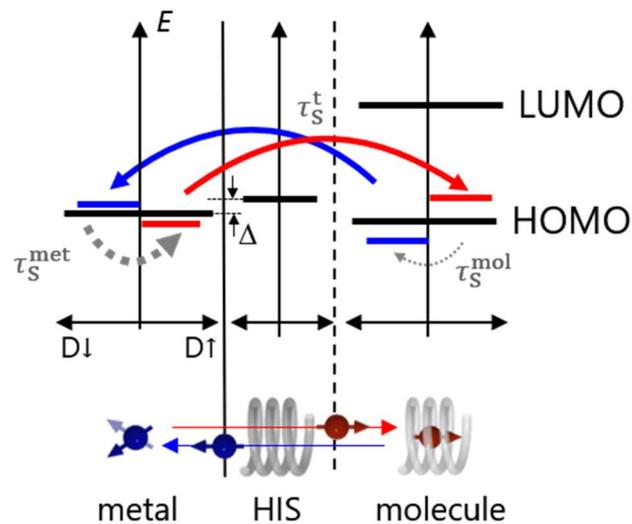

**Figure 4 Schematic image of thermally induced spin polarization.** Schematic illustration of thermally induced spin polarization due to charge transport at metal and chiral molecule inteface. Red and blue bending arrows denote thermally excited electron transport from metal to molecule and molecule to metal, respectively. i.e. thermerlly excited spin selective transport. Blue and red lines represent the spin-up DOS D $\downarrow$ and the spin-down DOS D $\uparrow$. $\tau_S^{met(mol)}$ and $\tau_S^t$ are spin relaxation time in metal(molecule) and transport time between metal and molecule.

**Acknowledgements**

We thank E. Minamitani, A. Shitade and M. Haze for fruitful discussions. This work was supported by JSPS-KAKENHI (18H03880, 19H02586).